\documentclass[amsmath,amssymb,superscriptaddress,showpacs]{revtex4}

\usepackage{graphicx}
\begin{document}

\title{Reply to the "Comment on 'Intrinsic tunneling spectra of $\rm Bi_2(Sr_{2-x}La_x)CuO_6$' ": Auxiliary information.} 

\author{A.Yurgens}
\affiliation{Chalmers University of Technology, 
G\"oteborg, Sweden} 

\author{D. Winkler}
\affiliation{Chalmers University of Technology, 
G\"oteborg, Sweden}
\affiliation{IMEGO Institute, Aschebergsgatan 46,
G\"oteborg, Sweden} 

\author{T. Claeson}
\affiliation{Chalmers University of Technology, 
G\"oteborg, Sweden}

\author{S. Ono} 
\affiliation{Central Research Institute of Electric Power Industry (CRIEPI),
Komae, Tokyo 201-8511, Japan}  

\author{Yoichi Ando} 
\affiliation{Central Research Institute of Electric Power Industry (CRIEPI),
Komae, Tokyo 201-8511, Japan} 

\noindent \today


\maketitle

\section{Introduction}
A poor thermal conductivity of all high-temperature superconductors makes them prone to local overheating whenever high current density is applied to them. Several works used short-pulse techniques to minimize effect of heating in stacks of intrinsic tunneling junctions (mesas) (ITJ)~\cite{1,2,3,4,5,6}. There were few attempts to measure directly the temperature of the stack utilizing the temperature-dependent c-axis resistance~\cite{7,8}. In such experiments, an extra stack placed on top of- ~\cite{7} or next to~\cite{8} the major one served as thermometer. Such a thermometer could not yet correctly measure temperature under the bias lead situated at relatively large distance from the thermometer. Due to larger current density in the area of bias contact, especially if there were many "extra" junctions underneath it because of technical problems~\cite{7}, such contacts is the major source of heating and the temperature there may be even higher than deduced in Refs.~\cite{7,8}.

On the other hand, estimations of overheating, based on a simple theory and numerical simulations  showed that the self-heating is proportional to a characteristic in-plane size of the stack~\cite{9}. Experimental data for small Bi2212 mesas were in qualitative agreement with the calculations. Estimations of self-heating in Bi2212 mesas with different sizes demonstrated that self-heating can effectively be obviated in small mesa structures~\cite{9}. 

To better address the heating issue, we directly measure temperature of the stack by using a micron-sized thermocouple which is in direct thermal- and electrical contact to the stack, exactly in the place where the bias current is injected. We believe that the temperature measured in this geometry is closest to the real one inside the stack.    

\section{Samples and measurements}
We have performed simultaneous temperature- and I-V measurements on a number of Bi-based single crystals, both 2201 and 2212, with three different sizes of stacks on every crystal,  4x4, 6x6, and  8x8 $\mu $m$^2$ in area, and containing from 15 to 30 ITJ, i.e. about 200-400~\AA \ high.  Those were wet-etched (in saturated solution of ethylenediaminetetraacetic acid (EDTA))~\cite{10}.  

\begin{center} 
\begin{figure}[th]
\includegraphics[width=8cm]{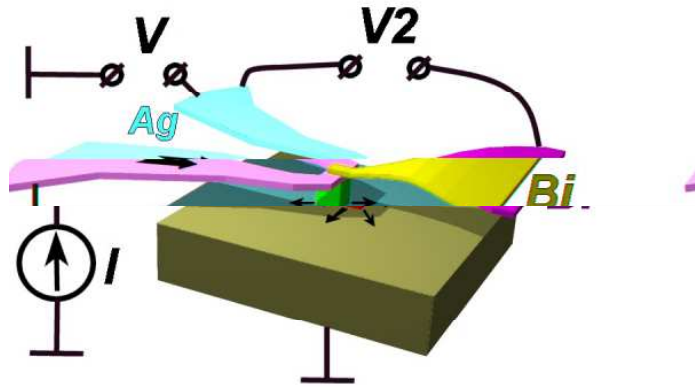}
\caption{Schematic layout of simultaneous temperature- and I-V- measurements.  Silver thin- film contacts are used for current (I) and voltage (V), while the bismuth one is used to monitor the temperature of the stack (red) by measuring voltage V2.  Black arrows represent current flow. } 
\label{Fig1}
\end{figure} 
\end{center} 

\begin{center} 
\begin{figure}[bh]
\includegraphics[width=8cm]{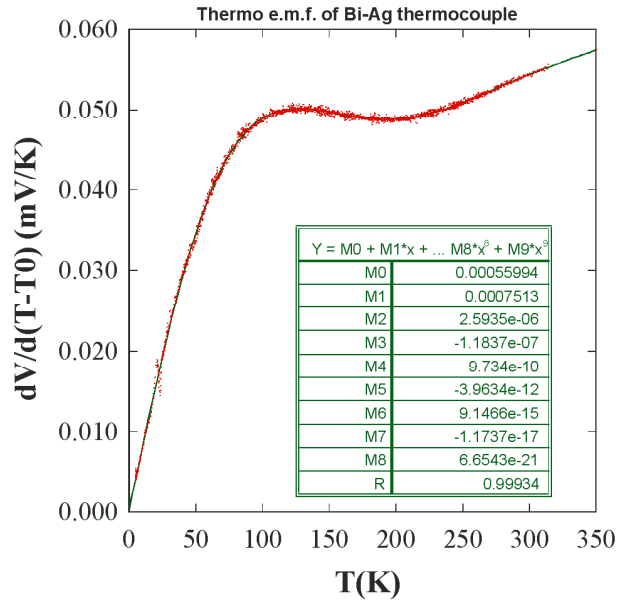}
\caption{ Thermo-e.m.f of  Bi-Ag thermocouple used in the experiments.  Table shows coefficients of polynomial fit through the experimental points.} 
\label{Fig2}
\end{figure} 
\end{center} 

\setcounter{footnote}{0}

Fig.~1 schematically shows a layout of contacts in the vicinity of stack (red). All the electrical contacts were patterned from metal thin films using photolithography and wet- and dry etching. The current ("I") and potential ("V") leads (light-blue) are formed by just one continuous silver thin film (200~nm thick). A 200-nm-thick bismuth thin film (magenta) is making a counter- electrode of the Ag-Bi thermocouple.  Simultaneously with the ones deposited onto the BSCCO-stacks, the silver and bismuth thin films were deposited on a separate glass substrate \footnotemark[\value{footnote}]\dag. The latter formed an equivalent thermocouple which was calibrated using small diode thermometers, in a separate experiment.  Rather large thermo-e.m.f. of such a thermocouple, about 50~$\mu$V/K at $100-300$~K (see Fig.~2), assures high accuracy of temperature measurements.  

\begin{center} 
\begin{figure}[t]
\includegraphics[width=8cm]{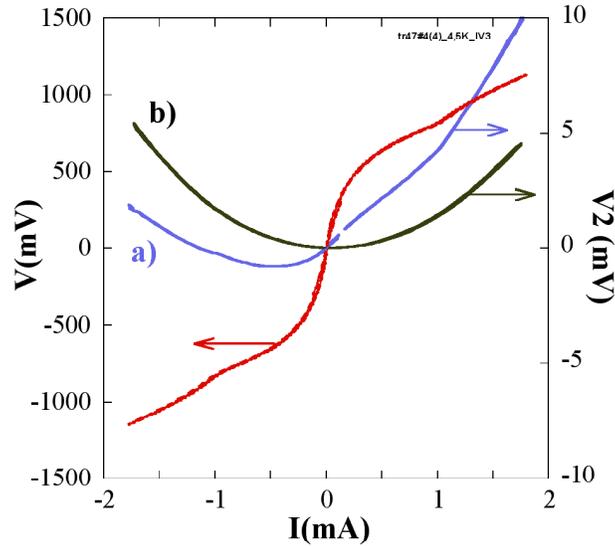}
\caption{ $V(I)$ and $V2(I)$ for two configurations of potential and current leads in the case when the Bi-contact was not accurately aligned. a) V- and I- contacts as scetched in Fig.~1; b) - with V- and I- contacts swapped.} 
\label{Fig3}
\end{figure} 
\end{center}

Due to much larger voltages $V\sim 1$~V appearing on the potential Ag-contact, Bi-contact with $V2 < 10$~meV is important to accurately align relative to the stack and other contacts. Misalignment results in picking some voltage drop (very small compared with $V$, but essential compared to $V2$) along the current lead leading to highly asymmetric $V2(I)$-dependence (Fig.~3a).  When the current and potential contacts were swapped, $V2$($I$ or $V$) became much more symmetric, Fig.~3b, implying that the Bi-contact was originally slightly shifted towards I-contact.  This allowed us to perform additional check to make sure that we measure $T$ exactly on the top of the stack.  From the symmetry of $V2(I)$ curves and their independence on configuration of voltage and current leads for a given stack, it is possible to say whether the Bi-contact was placed correctly.  
Measurements of all voltages and currents were done with help of instrumentation (pre)-amplifiers AMP01 (Fabr Analog Devices, input noise 5~nV/$\sqrt{\mathrm Hz}$) and 16-bits-20~kHz data acquisition card (PCI-16MIO from National Instruments) equipped with sample-and-hold input stage for simultaneous sampling of data.  Normally, a slow bias sweep (0.05~Hz) was used to measure both the temperature and I-V's and somewhat faster (0.5~Hz) sweep for just I-V's, at the same ambient temperature.  Totally, $2000-3000$ data points are normally logged in the former-, and 10000 in the latter case.  The calibration curve shown in Fig.~2 was used to calculate $T$ at each $I-V$ point. \begin{center} 
\begin{figure}[th]
\includegraphics[width=16cm]{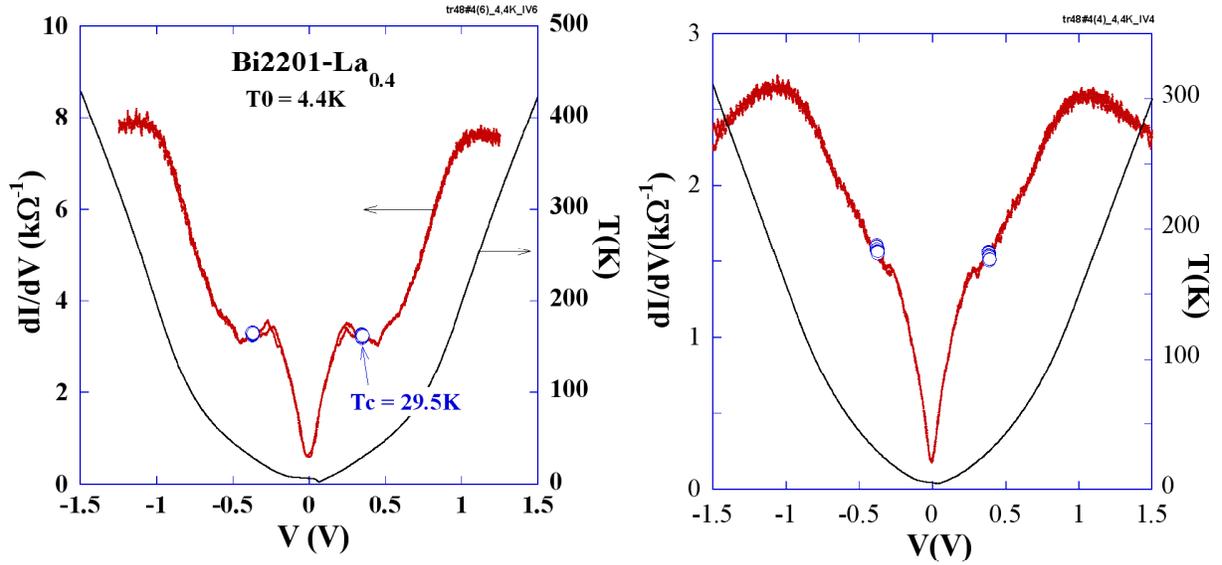}
\caption{ $dI/dV(V)$ and temperature of the Bi2201-La$_0.4$ (optimal doping) stacks.  Blue empty dots correspond to $T_s  = T_c$.  Left: 6x6-, right: 4x4 $\mu$m$^2$  in area;} 
\label{Fig4}
\end{figure} 
\end{center}   

\section{Results}
In all cases, the temperature of the stack $T_s$ can easily rise as high as $200-300$~K above the ambient temperature $T_0$, see Figs.~4-6.  This means that the humps which are seen in intrinsic tunneling spectra at high bias, are much likely due to overheating, as correctly stated in the Comment by V.N. Zavaritsky~\cite{11,12}. 

\setcounter{footnote}{0}

Fig.~4  shows $dI/dV(V)$ and temperature of the Bi2201-La$_{0.4}$ (optimal doping) stacks~\footnotemark[\value{footnote}]\ddag.  The former curves show all characteristic features of a typical intrinsic-tunneling spectrum.  There are superconducting-gap (SG) peaks, then, at higher voltages - "dips"~\cite{13}, eventually followed by "humps". The latter two were usually attributed to the pseudogap (PG)~\cite{13,14,15,16,17}. 
 
\begin{center} 
\begin{figure}[th]
\includegraphics[width=8cm]{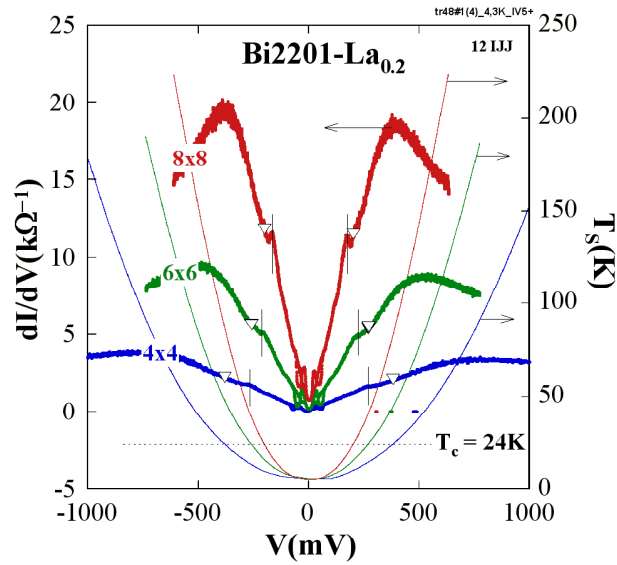}
\caption{ $dI/dV(V)$ and $T_s(V)$ for three stacks of different areas, 8x8 (red), 6x6 (green), and 4x4 $\mu$m$^2$ (blue) prepared on one and the same single crystal of Bi2201-La$_{0.2}$ (overdoped).  All stacks had 12 IJJ.  Triangular marks indicate the voltages at which  $T_s \approx T_c \approx 24$~K.  Vertical black lines indicate the positions of superconducting peaks (although not well seen for this sample).  For "8x8" stack, they correspond to $\approx 18-19$~K, for "6x6" - $\approx 17-18$~K, and for "4x4" - $\approx 14$~K, confirming that for small sizes the overheating is smaller~\cite{9}.} 
\label{Fig5}
\end{figure} 
\end{center} 
 
Our temperature measurements have shown that the SG-peaks develop at voltages, corresponding to higher than ambient temperature (yet at $T < T_c$), that agrees with what was qualitatively shown earlier~\cite{1,2,3,4,5,6,7,8}. However, the PG humps always lie at $T > T_c$, see Figs.~4 and 5. Big empty marks indicate voltages at which  $T_s \approx T_c$.  In Fig.~5, the vertical black lines indicate positions of the SG peaks which are not well resolved in the plot.  For "8x8" stack the lines correspond to $\approx 18-19$~K, for "6x6" - $\approx 17-18$~K, and for "4x4" - $\approx 14$~K, confirming that for small sizes the overheating is smaller~\cite{9}. The SG is accordingly smaller at higher temperature that explains the difference in SG's between samples with the same number of IJJ.
\begin{center} 
\begin{figure}[bh]
\includegraphics[width=8cm]{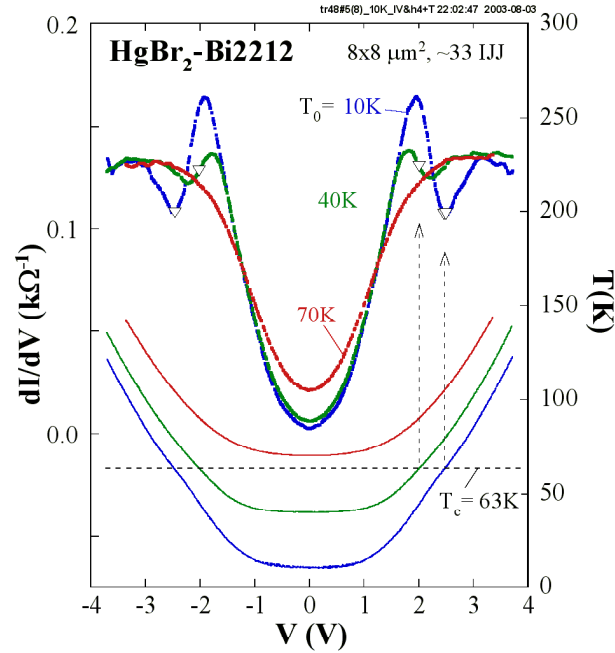}
\caption{$dI/dV(V)$ and temperature of the HgBr$_2$-Bi2212 8x8 $\mu$m$^2$ stack. Empty triangles correspond to $T_s=T_c\approx 63$~K. Note flat parts of $T_s(V)$-curves at $-1<V<1$~Volts. Within this range, the overheating is less than 5-6~K, see Fig.~7.} 
\label{Fig6}
\end{figure} 
\end{center}

To reduce internal heating in stacks of IJJ, one can use the so called intercalated samples, where relatively big chemically neutral HgBr$_2$-molecules are interspersed inside Bi2212 single crystals~\cite{18}. The intercalated molecules were found to locate themselves in between adjacent BiO layers, thus pulling apart superconducting CuO$_2$-layers.  This results in dramatic reduction of the $c$-axis critical current density (typically 10-30 times) and increase of the tunneling resistance, and in very much anticipated decrease of the internal heating~\cite{14}. These intercalated samples allowed us to reach the normal-state tunneling parts without visible signatures of internal heating (back-bending). 

\begin{center} 
\begin{figure}[th]
\includegraphics[width=8cm]{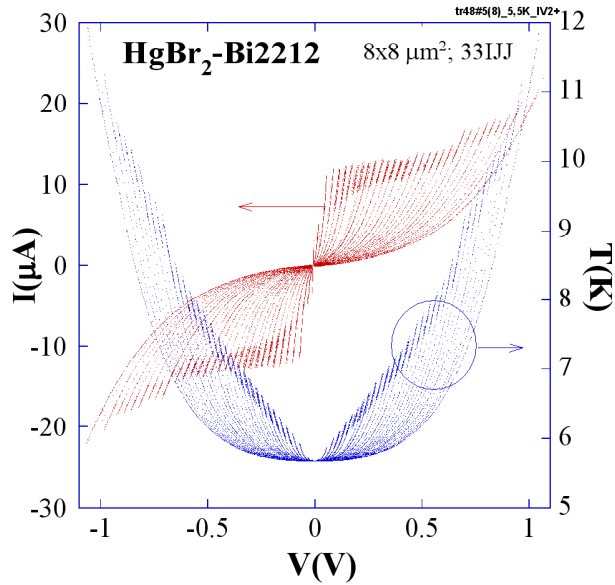}
\caption{$I(V)$-curves and temperature of the HgBr$_2$-Bi2212  8x8 $\mu$m$^2$ stack, the same as in Fig.~6. The $T_s(V)$ multi-branch curves were calculated from the (experimentally measured) average $T_s(Q)$-dependence, where $Q$ is the dissipated power.  Note that, despite  a large number of IJJ in this particular stack, the overheating does not exceed $5-6$~K within ±1 V of bias voltage.} 
\label{Fig7}
\end{figure} 
\end{center}   

However, from Fig.~6 we see that internal heating is still a problem at relatively high bias even in those samples.   $T_s \approx T_c$ at voltages which correspond to the positions of the dips.
 
\begin{center} 
\begin{figure}[bh]
\includegraphics[width=8cm]{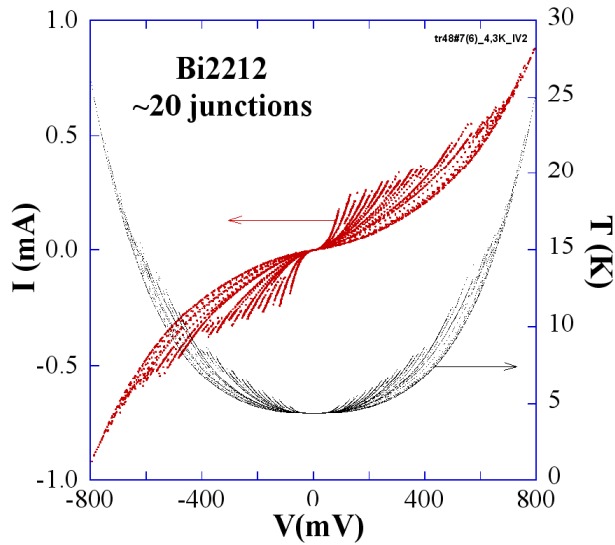}
\caption{Multi-branch hysteretic $I(V)$-curves and temperature of the Bi2212  6x6 $\mu$m$^2$ stack.  For  this particular stack, the overheating does not exceed 20~K within $\pm1$~V of bias voltage.} 
\label{Fig8}
\end{figure} 
\end{center}  
 
Overheating at biases corresponding to the hysteretic multi-branch $I(V)$ characteristics is nevertheless relatively small.  For the intercalated sample, $\delta T < 5-6$~K at most, despite the number of junctions is big (33, compared to a "normal" for our experiments number of 10-15).  For the pristine Bi2212- single crystals, this overheating is about 4-5 times larger, see Fig.~8. 

It is important to take into account much likely different thermal conductivities of pristine and the intercalated samples, while comparing the results for them. It was noted, that the heating rate, $dT/dQ$, is different, $dT/dQ \sim 30$~K/mW for the pristine, and $\sim $70~K/mW for the intercalated sample.  $dT/dQ$ was found to be almost constant for high temperatures and has tendency to somewhat increase on cooling.

\setcounter{footnote}{0}
\begin{center} 
\begin{figure}[th]
\includegraphics[width=16cm]{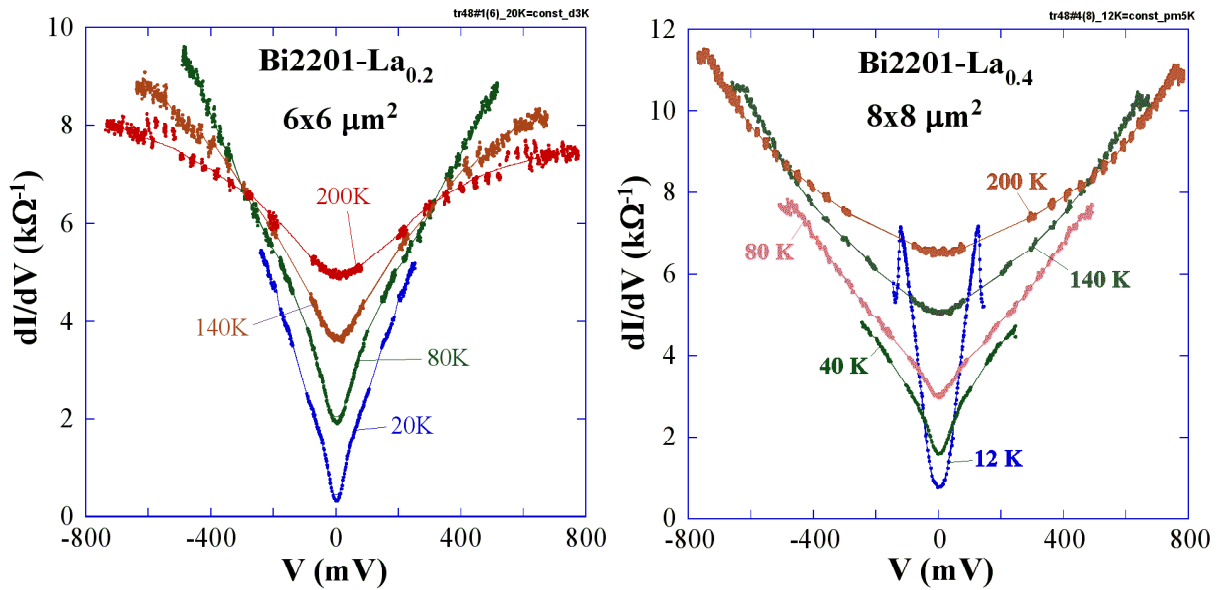}
\caption{Bi2201-La$_x$. The "heating-free"  tunneling-conductance curves deduced from the "pool" of numerous "3D" $I-V-T_s$ data points taken at about $30-40$ bath temperatures in the range $4-220$~K.  The $\sigma(V)= dI/dV(V)$-points in these plots correspond to temperatures only different to within $\pm 1.5$~K from the temperatures shown along with the curves (with one exception for $T_{s0}= 12$~K to the right, for which $\delta T = \pm 2.5$~K).  Note also that $\sigma(V) \propto |V|^{\alpha}$, with $\alpha\leq 1$ for low temperatures.  Lines are guides to the eye.}
\label{Fig9}
\end{figure} 
\end{center} 

\begin{center} 
\begin{figure}[th]
\includegraphics[width=8cm]{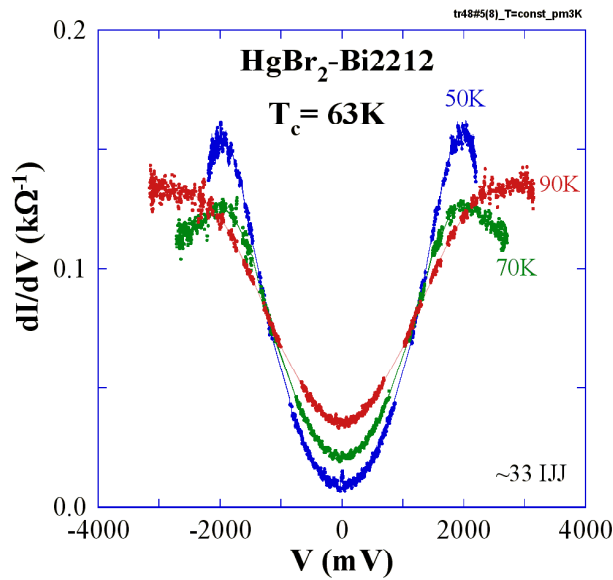}
\caption{ HgBr$_2$-Bi2212 stacks. The "heating-free"  tunneling-conductance curves deduced out from the pool of numerous "3D" $I-V-T_s$ data points taken at about 30 bath temperatures in the range $4-90$~K.  The $\sigma(V)= dI/dV(V)$-points in these plots correspond to temperatures only different to within $\pm 1.5$~K from the temperatures shown along with the curves.   Note that $\sigma(V)(V)\propto |V|^{\alpha}$, with ${\alpha} > 1$ for all temperatures, in contrast to Bi2201-La$_x$ stacks, for which ${\alpha}\leq 1$, see Fig.~9. Lines are guides to the eye. } 
\label{Fig10}
\end{figure} 
\end{center}    

The detailed measurements of the temperature of the stacks at different ambient temperatures $T_0$ allow us to deduce the unique, "heating-free" $I(V)$-, or $dI/dV(V)$-curves.  First, we put the measurements for a given sample ($I-V-T_s$-triples) at all $T_0$,  into a single file.  Then, we pick out those $I-V$ pairs which correspond to a specified $T_{s0}$ and $\delta T$, i.e. "allowed" range of overheating.  In the following, we choose $\delta T = \pm1.5$~K, counted from the chosen $T_{s0}$.
Figs.~9 and 10 show results of such a selection.  

Figs.~9 and 10 show that, after the heating is properly taken into account, the resulting tunneling conductance curves are of very much the same non-linear character, except for the either absence of the humps, or their being located at much higher than working-range voltages~\footnotemark[\value{footnote}]\S.  One can see some tendency for developing the humps in Fig.~9 (right), and in Fig.~10.  Unfortunately, for the stacks we measured, the heating rate was too high to reach suggestive voltages presumably corresponding to humps without significant overheating.  Perhaps, stacks with linear in-plane sizes in the sub-$\mu $m range are needed to finally elucidate this issue.

\section{Discussion}
The previous section in its last part clearly shows that the calculations of the Comment~\cite{11,12}, although being valid with respect to the general idea of  self-heating in the stacks of IJJ, fail to be quantitatively correct.  This is because the Comment assumed that the resistance is bias independent~\cite{11,12}, a simplification which is certainly incorrect, see Figs.~9 and 10. It is clear, for instance, that within Newton's cooling scenario, no form of suggested bias-independent $R_c(T)$ can account for the observed  $\sigma (V)\propto |V|^{\alpha }, \alpha<1$, (which is further supported by break- and point-contact- junction experiments), ~\cite{19,20,21}.  One has to take into account the highly non-linear $I(V)$-curves of Fig.~9 or 10 to fit better experimental results with help of suggested Newton's cooling scenario.

Anyway, the self-overheating turns out to be a severe problem in the intrinsic tunneling spectroscopy of superconducting gap and pseudogap, mainly due to a very bad thermal conductivity of the host material and large values of the gaps involved.

This problem, although not obviously, can also be acute in some low-resistance break-junction experiments. Indeed, the break-junction contact in high-$T_c$-related experiments is, in overwhelming majority of such experiments, of unknown shape, area and location inside the break.  Thermal-conductivity properties of such a contact are therefore unknown, too.  It may happen, that the contact is located at a distance from, say, a metal needle used to produce the break~\cite{22}. Then, the only escape path for the developed Joule heat goes through the poorly conducting Bi2212, i.e. one has the same prerequisites for the overheating as in the intrinsic-tunneling experiments.  

Let us make an estimation of the overheating for the (almost) worst case, when the heat is only removed by the thermally conducting helium gas. We will ignore somewhat better conducting Bi2212 single crystals, leads, gas convection, etc.  Then, in the spherical geometry, and taking into account the temperature dependence of the thermal conductivity of He~\cite{23}, which can be approximated by a simple power-low dependence: $\kappa(T) \approx 0.153\ (T/300)^{0.656} [\mathrm{W/m\ K}]$, we have an approximate estimation: 
$\delta T [ \mathrm{K} ] \sim 140\cdot (Q [ \mathrm{mW} ] / r [ \mathrm{\mu m} ])^{0.60}$, i.e. 140~K for $Q=2$~mW and the linear sizes $r$ of about 2~$\mu $m. 

This formula gives an-order-of-magnitude-correct estimate of the overheating in the case of intrinsic junctions.  It gives anything between 100 and 1000 K for break-junctions, if we take the range of junction resistances $10-3000\Omega$~\cite{24} and characteristic voltages $\sim $200~mV, and assume that the tunneling junction size is of the order of  $100-1000$~\AA. The latter is most uncertain, but it is very hard to expect that a torn break can have flat and parallel tunneling-junction surfaces of linear size larger than 0.1~$\mu $m, unless it is made of fragments of the intrinsic junctions themselves.

\section{Conclusions}
In conclusion, we have measured directly local overheating in the stacks of intrinsic tunneling junctions exactly at places where the bias current is injected.  Different Bi-based materials, single-layer, double-layer, and intercalated double-layer ones were investigated.  The measurements were motivated by the recent Comment, ~\cite{11,12} which, by using simple Newton's cooling scenario, was able to reproduce high-bias features (pseudogap humps) of our recent work~\cite{17}, thus doubting the validity of those.  We confirm the main conclusion of the Comment, that the "pseudogap features" reported in our paper are an artifact of Joule heating.

However, our measurements have shown that, despite the severe heating in the stacks, there is yet a way of extracting the genuine isothermal I-V's.  Such $I(V,T=\mathrm{const})$ appear to be highly non-linear in the whole temperature range and show notable differences with regard to single- or double-layer compounds.  It is also concluded that at moderate biases, where the hysteretic multi-branch I-V's are seen, the heating can be very small, on the level of few Kelvin.

Comparison of intrinsic-junction geometry with other tunneling-spectroscopy techniques, like break-junction ones, has revealed, that the heating issues can be acute there as well.  All this means that results of tunneling spectroscopy on high-$T_c$ materials should be taken with discretion.

\acknowledgements{Discussions with V. Krasnov are appreciated.}

\footnotetext[\value{footnote}]{\dag It seems to be important to deposit Bi thin film used for calibration and the main thin film in one and the same run.  Bi-Ag thermocouple deposited in another run apparently had ~30\% lower thermo-e.m.f. }      

\footnotetext[\value{footnote}]{\ddag The number of IJJ is only approximately known  to be about $7-10$ for that sample.  The uncertainty is because the branches were not well seen. When the individual "gaps" are not equal (say, due to different doping levels), I-V's show rather chaotic switching between different combinations of "good", and "bad" junctions that makes counting them difficult.} 

\footnotetext[\value{footnote}]{\S There are no minima in $R_c(T<300\ \mathrm{K})$ for the samples shown in Fig.~9. This is noted to be inconsistent with their nominal La-content of 0.2 (overdoped material, left) and 0.4 (optimal doping, right). The samples may have lost some oxygen during fabrication of the stacks. It is therefore not possible to conclude whether the non-linearities of the heating-free $dI/dV(V)$'s shown in Figs.~9 should vanish at the positions of minima in the corresponding $R_c(T)$'s~\cite{17}.}

\end{document}